\documentclass[10pt,twocolumn,twoside]{IEEEtran}
\usepackage{graphicx}
\usepackage{amsmath}

\usepackage{amsfonts}
\usepackage{amssymb}
\usepackage{array}
\newcommand{\PreserveBackslash}[1]{\let\temp=\\#1\let\\=\temp}
\newcolumntype{C}[1]{>{\PreserveBackslash\centering}p{#1}}
\newcolumntype{R}[1]{>{\PreserveBackslash\raggedleft}p{#1}}
\newcolumntype{L}[1]{>{\PreserveBackslash\raggedright}p{#1}}
\usepackage[usenames]{color}
\usepackage{colortbl,booktabs}
\usepackage{tabularx}
\usepackage{multicol}
\usepackage{booktabs}
\usepackage[Symbol]{upgreek}
\usepackage{subfigure}
\usepackage{bm}
\usepackage{cite}
\usepackage{balance}
\usepackage[ruled,vlined]{algorithm2e}

\usepackage{threeparttable}
\usepackage{amsmath}
\usepackage{dcolumn}
\usepackage{multirow}
\usepackage{amsfonts}
\newcolumntype{d}[1]{D{.}{.}{#1}}

\begin{document}

\bibliographystyle{IEEEtran} 
\title{Channel Estimation for Extremely Large-Scale Massive MIMO: Far-Field, Near-Field, or Hybrid-Field?}

\author{Xiuhong Wei and Linglong Dai
\vspace{-2em}
\thanks{All authors are with the Beijing National Research Center for Information Science and Technology (BNRist) as well as the Department of Electronic Engineering, Tsinghua University, Beijing 100084, China (e-mails: weixh19@mails.tsinghua.edu.cn, daill@tsinghua.edu.cn).}
\thanks{This work was supported in part by the National Key Research and Development Program of China (Grant No. 2020YFB1807201) and in part by the National Natural Science Foundation of China (Grant No. 62031019).}
}

\maketitle
\vspace{-5mm}
\begin{abstract}
Extremely large-scale massive MIMO (XL-MIMO) is a promising technique for future 6G communications. However, existing far-field or near-field channel model mismatches the hybrid-field channel feature in the practical XL-MIMO system. Thus, existing far-field and near-field channel estimation schemes cannot be directly used to accurately estimate the hybrid-field XL-MIMO channel. To solve this problem, we propose an efficient hybrid-field channel estimation scheme by accurately modeling the XL-MIMO channel. Specifically, we firstly reveal the hybrid-field channel feature of the XL-MIMO channel, where different scatters may be in far-field or near-field region. Then, we propose a hybrid-field channel model to capture this feature, which contains both the far-field and near-field path components. Finally, we propose a hybrid-field channel estimation scheme, where the far-field and near-field path components are respectively estimated. Simulation results show that the proposed scheme performs better than existing schemes.
\end{abstract}


\section{Introduction}\label{S1}

With the emergence of new applications, 6G is expected to achieve a 10-fold increase in spectrum efficiency than 5G~\cite{6G}. The extremely large-scale massive MIMO (XL-MIMO) is a promising technique for 6G to achieve this goal, where the base station (BS) deploys an extremely large number of antennas to achieve higher spectral efficiency and improved energy efficiency~\cite{XLMIMO}. However, the sharp increase of BS antennas leads to the unaffordable pilot overhead for the high-dimensional XL-MIMO channel estimation.

There are two typical categories of low-overhead channel estimation schemes for XL-MIMO, i.e., far-field channel estimation~\cite{OMP,XinyuGao_beamsqilt_TSP,My} and near-field channel estimation~\cite{Mingyao,JinShi}. The first category is the {\it{far-field}} channel estimation by considering the channel sparsity in the {\it{angle domain}}. In this category of schemes, the XL-MIMO channel is modeled in the far-field region with the planar wave assumption. Based on this assumption, the array steering vector of the channel is only related to the angle. With the help of the classical discrete fourier transform (DFT) matrix, the non-sparse spatial channel can be firstly represented by the sparse angle-domain channel. Then, some compressive sensing (CS) algorithms such as orthogonal matching pursuit (OMP)~\cite{OMP} can be used to estimate this sparse angle-domain channel with low pilot overhead. 

The second category is the {\it{near-field}} channel estimation by considering the channel sparsity in the {\it{polar domain}}. Specifically, since the array aperture of XL-MIMO is very large, the XL-MIMO channel can be more accurately modeled in the near-field region with the spherical wave assumption. Under this assumption, the array steering vector of the channel is not only related to the angle, but also related to the distance between the BS and the scatter~\cite{Near}. Based on the near-field channel model, a few near-field channel estimation schemes have been proposed recently~\cite{Mingyao,JinShi}. Specifically,  a new polar-domain sparse representation of the original XL-MIMO channel in the spatial domain was proposed~\cite{Mingyao}, where the transform matrix was generated from the joint angle and distance space to replace the classical DFT matrix only associated with the angle space. By considering this channel sparsity in the polar domain, the corresponding CS algorithms have been proposed to reduce the pilot overhead for the near-field channel estimation~\cite{Mingyao, JinShi}.


In the existing far-field or near-field channel model, it is assumed that all scatters are either in the far-field or near-field region. Actually, a hybrid-field communication environment is more likely to appear in the XL-MIMO system, where some scatters are in the far-field region, while others may locate in the near-field region. In other words, the XL-MIMO channel is usually composed of both the far-field and near-field path components. However, the existing far-field or near-field channel model mismatches this hybrid-field channel feature, which makes the existing far-field or near-field channel estimation schemes cannot be directly used to accurately estimate the hybrid-field XL-MIMO channel. Unfortunately, this important problem has not been studied in the literature.

To fill in this gap, we propose an efficient hybrid-field channel estimation scheme by accurately modeling the hybrid-field XL-MIMO channel in this paper{\footnote{\leftskip=0pt \rightskip=0pt plus 0cm{Simulation codes are provided in the following link to reproduce the results\\ presented in this paper:  http://oa.ee.tsinghua.edu.cn/dailinglong/ publications/\\publications.html.}}}. Our contributions are summarized as follows.

\begin{enumerate}

\item We reveal the hybrid-field channel feature of the XL-MIMO channel. Specifically, this hybrid-field channl feature means that different scatters may be in different regions. On the one hand, some scatters are far away from the BS, which are in the far-field region. On the other hand, some scatters are relatively close to the BS, which are in the near-field region.

\item In order to capture this hybrid-field channel feature, we propose a hybrid-field channel model for the XL-MIMO channel. In the proposed hybrid-field channel model, both the far-field and near-field path components in the far-field and near-field regions are considered, which correspond to different sparse representations based on different channel transform matrices. Moreover, we can control the proportion of the two types of path components by using an adjustable parameter. In this way, the existing far-field and near-field channel models can be regarded as special cases of the proposed hybrid-field channel model.

\item Based on this hybrid-field channel model, we propose a hybrid-field channel estimation scheme to estimate the XL-MIMO channel with low pilot overhead. The basic idea is to individually estimate the far-field and near-field path components by using different channel transform matrices. The far-field path components are estimated by considering their sparsity in the angle domain, while the near-field path components are estimated by considering their sparsity in the polar domain. Particularly, the existing far-field and near-field channel estimation schemes can be regarded as special cases of the proposed scheme.

\end{enumerate}

The rest of the paper is organized as follows. In Section II, we introduce the signal model, and review two existing far-field and near-field channel models. In Section III, we firstly reveal the hybrid-field channel feature, and then propose the hybrid-field channel model and the corresponding hybrid-field channel estimation scheme. Simulation results and conclusions are provided in Section IV and Section V, respectively.

{\it Notation}: Lower-case and upper-case boldface letters ${\bf{a}}$ and ${\bf{A}}$ denote a vector and a matrix, respectively; ${{{\bf{a}}^H}}$ and ${{{\bf{A}}^{H}}}$ denote the conjugate transpose of vector $\bf{a}$ and matrix $\bf{A}$, respectively; ${{\|{\bf{a}}\|_2}}$ denotes the $l_2$ norm of vector $\bf{a}$. Finally, $\cal CN\left(\mu,\sigma \right)$ denotes the probability density function of the circularly symmetric complex Gaussian distribution with mean $\mu$ and variance $\sigma^2$,  and ${{\cal U}(-a,a)}$ denotes the probability density function of uniform distribution on $(-a,a)$.
\vspace{-2mm}

\section{System Model}\label{S2}
In this section, we will first introduce the signal model of the XL-MIMO system. Then, the existing far-field and near-field channel models will be briefly reviewed, respectively.


\vspace{-2mm}
\subsection{Signal Model}\label{S2.1}
We consider that the BS employ a $N$-element extremely large-scale antenna array to communicate with a single-antenna user. Let ${\bf{h}}^H\in\mathbb{C}^{1\times N}$ denote the channel from the BS to the user. Take the downlink channel estimation as an example, the corresponding signal model can be represented by
\begin{equation}\label{eq1}
{\bf{y}}^H = {\bf{h}}^H{\bf{P}}^H + {\bf{n}}^H,
\end{equation}
where ${{\bf{y}}^H}\in\mathbb{C}^{1\times M}$ represents the received pilots by the user in $M$ time slots, ${{\bf{P}}^H}\in\mathbb{C}^{N\times M}$ represents the transmitted pilot signals by the BS in $M$ times slots, and ${{\bf{n}}}\sim{\cal C}{\cal N}\left( {0,\sigma^2{\bf{I}}_{M}} \right)$ represents the ${{M} \times 1}$ received noise in $M$ times slots with ${\sigma^2}$ representing the noise power.

Through the conjugate transpose transformation,~(\ref{eq1}) can be further formulated as
\begin{equation}\label{eq2}
{\bf{y}} = {\bf{Ph}} + {\bf{n}}.
\end{equation}
The downlink channel estimation is to estimate $\bf{h}$ on the premise that ${\bf{y}}$ and ${\bf{P}}$ are known. In the XL-MIMO system, the number of antennas $N$ at the BS is large. In order to reduce the pilot overhead, the low-overhead channel estimation should be investigated so that the number of pilots $M$ is much smaller than $N$. Next, we will briefly review two existing channel models for existing channel estimation schemes.

\vspace{-3mm}
\subsection{Channel Models}\label{S2.2}

Specifically, as shown in Fig. 1, the electromagnetic radiation field in wireless communication systems can be divided into far-field and near-field~\cite{Mingyao}, where different fields will result in different channel models. The bound between these two fields is determined by the Rayleigh distance $Z=\frac{2D^2}{\lambda}$, where $D$ and $\lambda$ are the array aperture and wavelength, respectively.

\vspace{-3mm}
\begin{figure}[htbp]
\begin{center}
\includegraphics[width=0.8\linewidth]{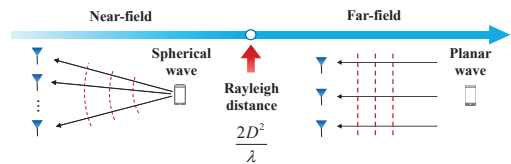}
\end{center}
\setlength{\abovecaptionskip}{-0.3cm}
\caption{The near-field region and the far-field region~\cite{Mingyao}.}. \label{HF}
\end{figure}

\subsubsection{Far-Field Channel Model}\label{S2.2.1}

As shown in Fig. 1, when the distance between the BS and the scatter is larger than the Rayleigh distance, the far-field channel ${\bf{h}}_{\rm{far\mbox{-}field}}$ is modeled under the planar wave assumption, which can be represented by
\begin{equation}\label{eq3}
{\bf{h}}_{\rm{far\mbox{-}field}}=\sqrt{\frac{N}{L}}\sum\limits_{l = 1}^{L}{\alpha_{l}}{\bf{a}}\left({\theta}_l\right),
\end{equation}
where $L$ represents the number of path components (corresponding to effective scatters) between the BS and the user, ${\alpha}_{l}$ and ${\theta}_l$ represent the gain and angle for the $l$th path, respectively. ${\bf{a}}\left({\theta}_l\right)$ represents the far-field array steering vector based on the planar wave assumption, which can represented by
\begin{equation}\label{eq4}
{\bf{a}}\left( {\theta_l}\right) = \frac{1}{{\sqrt N }}{\left[1,{{e^{ - j{\pi}{\theta_l}}}},\cdots,{{e^{ - j(N-1){\pi}{\theta_l}}}} \right]^H}.
\end{equation}
It is noted that ${\theta}_l=2\frac{d}{\lambda}{\rm{cos}}(\phi_l)$, where $d={\frac{\lambda}{2}}$ is the antenna spacing, and  $\phi_l\in(0,\pi)$ is the practical physical angle.

In order to reduce the pilot overhead for the channel estimation, the above non-sparse channel ${\bf{h}}_{\rm{far\mbox{-}field}}$ can be represented by the sparse angle-domain channel ${{\bf{h}}}^{{A}}_{\rm{far\mbox{-}field}}$ with the DFT matrix ${\bf{F}}$ as follows
\begin{equation}\label{eq5}
{{\bf{h}}}_{\rm{far\mbox{-}field}} = {\bf{F}}{{\bf{h}}}^{{A}}_{\rm{far\mbox{-}field}},
\end{equation}
where ${\bf{F}}=[{\bf{a}}\left( {\theta}_1\right),\cdots,{\bf{a}}\left( {\theta}_N\right)]$ is a $N\times N$ unitary matrix, where the columns are orthogonal to each other, and ${\theta}_n={\frac{2n-N-1}{N}}$ with $n=1,2,\cdots,N$. Since there are limited scatters in communication environments, the angle-domain channel ${{\bf{h}}}^{A}_{\rm{far\mbox{-}field}}$ is usually sparse. Based on this sparsity, some CS algorithms can be used to estimate this high-dimensional channel with low pilot overhead~\cite{OMP,XinyuGao_beamsqilt_TSP,My}.

\subsubsection{Near-Field Channel Model}\label{S2.2.1} In addition to the above far-field channel model, a near-field channel model was recently proposed in~\cite{Mingyao}. As shown in Fig. 1, when the distance between the BS and the scatter is smaller than the Rayleigh distance, the near-field channel is modeled under the spherical wave assumption, which can be represented by

\begin{equation}\label{eq6}
{\bf{h}}_{\rm{near\mbox{-}field}}=\sqrt{\frac{N}{L}}\sum\limits_{l = 1}^{L}{\alpha_{l}}{\bf{b}}\left({\theta}_l,r_l\right).
\end{equation}
Compared with the far-field channel model~(\ref{eq3}), the array steering vector ${\bf{b}}\left({\theta}_l,r_l\right)$ for the near-field channel model is derived based on the the spherical wave assumption, which can represented by~\cite{Mingyao}
\begin{equation}\label{eq7}
{\bf{b}}(\theta_l, r_l) = \frac{1}{\sqrt{N}}[e^{-j{\frac{2\pi}{\lambda}}(r_{l}^{(1)} - r_{l})},\cdots, e^{-j{\frac{2\pi}{\lambda}}(r_{l}^{(N)} - r_{l})}]^H,
\end{equation}
where $r_l$ represents the the distance from the $l$th scatter to the center of the antenna array, $r_{l}^{(n)} = \sqrt{r_l^2 + \delta_n^2d^2 - 2r_l\delta_n d\theta_l}$ represents the distance from the $l$th scatter to the $n$th BS antenna, and $\delta_n = \frac{2n - N - 1}{2}$ with $n = 1,2,\cdots, N$.

Since the DFT matrix in~(\ref{eq5}) associated with the angle domain only matches the array steering vector in the far-field, the near-field channel in~(\ref{eq6}) will cause serious energy spread in angle domain. In order to explore the sparsity of the near-field channel, a polar-domain transform matrix $\bf{W}$ was proposed in~\cite{Mingyao}, which can be represented by
\begin{equation}\label{eq8}
\begin{aligned}
{\bf{W}}=[{\bf{b}}(\theta_1, r_1^1),\cdots,{\bf{b}}(\theta_1, r_{1}^{S_{1}}),\cdots,\\{\bf{b}}(\theta_{N}, r_{N}^1),\cdots,{\bf{b}}(\theta_{N}, r_{N}^{S_N})],
\end{aligned}
\end{equation}
where each column of $\bf{W}$ is a near-field array steering vector with the sampled angle $\theta_{n}$ and distance $r_{n}^{s_n}$, with $s_n=1,2,\cdots,{S_n}$, $S_n$ represents the number of sampled distances at the sampled angle $\theta_{n}$. Thus, the number of all sampled grids can be represented by $S=\sum\limits_{n = 1}^{N}S_n$. Based on this polar-domain transform matrix $\bf{W}$, the near-field channel can be represented by
\begin{equation}\label{eq9}
{{\bf{h}}}_{\rm{near\mbox{-}field}} = {\bf{W}}{{\bf{h}}}^P_{\rm{near\mbox{-}field}},
\end{equation}
where ${{\bf{h}}}^P_{\rm{near\mbox{-}field}}$ is the $S\times 1$ polar-domain channel.  Similar to the far-field channel in the angle domain, ${{\bf{h}}}^{P}_{\rm{near\mbox{-}field}}$ also shows a certain sparsity. The corresponding polar-domain based CS algorithm was further proposed to reduce the pilot overhead for the near-field channel estimation~\cite{Mingyao}. However, since the polar transform matrix $\bf{W}$ is generated from the joint angle and distance space, its dimension is large, and the columns orthogonality is relatively poor compared with the  DFT matrix. Thus, the far-field channel in the polar domain may cause more serious energy leakage than that in the angle domain.

In all existing works above, all scatters in the communication environment are assumed to be either in the far-field or near-field region. In practical XL-MIMO communication environments, it is more likely that some scatters are in the far-field region, while others may locate in the near-field region. However, this hybrid-field communication environment cannot be accurately modeled by the existing far-field or near-field channel model, so existing far-field and near-field channel estimation schemes cannot be directly used to accurately estimate the hybrid-field XL-MIMO channel.

\section{Proposed Hybrid-Field Channel Estimation}\label{S3}
In this section, we will firstly reveal the hybrid-field feature for the XL-MIMO channel. Then, a hybrid-field channel model will be proposed to capture this channel feature. Finally, based on the proposed channel model, we will propose a hybrid-field channel estimation scheme to improve the estimation accuracy with low pilot overhead.

\vspace{-0.5em}
\subsection{Hybrid-Field Channel Feature}\label{S3.1}
\vspace{-1em}

\begin{figure}[htbp]
\begin{center}
\includegraphics[width=0.7\linewidth]{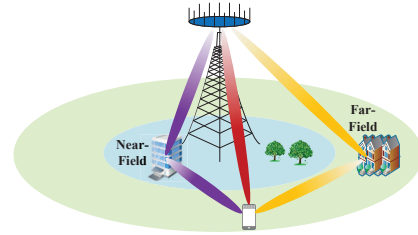}
\end{center}
\setlength{\abovecaptionskip}{-0.1cm}
\caption{The hybrid-field communication environment for XL-MIMO.} \label{RIS}
\end{figure}

As shown in Fig. 2, there are two different types of scatters for the XL-MIMO system. When the scatter is far away from the BS, it is in the far-field region of the BS, which will introduce the far-field path component. When the scatter is close to the BS, it is in the near-field of the BS, which will produce the near-field path component. For example, when a high-altitude BS serves a distant user, although the direct link produces a far-field path component, the scatters around the BS may generate the near-field path components~\cite{Mingyao}.

However, the existing far-field or near-field channel model is only composed of far-field or near-field path components, which cannot capture this hybrid-field feature of the XL-MIMO channel.

\subsection{Proposed Hybrid-Field Channel Model}\label{S3.1}

In order to capture this hybrid-field feature of the XL-MIMO channel mentioned above, we propose a hybrid-field channel model represented as
\begin{equation}\label{eq10}
\begin{aligned}
{\bf{h}}_{\rm{hybrid\mbox{-}field}}=\sqrt{\frac{N}{L}}\left(\sum\limits_{l_{\rm{f}} = 1}^{\gamma L}{\alpha_{l_{\rm{f}}}}{\bf{a}}\left({\theta}_{l_{\rm{f}}}\right)+\sum\limits_{l_{\rm{n}} = 1}^{(1-\gamma)L}{\alpha_{l_{\rm{n}}}}{\bf{b}}\left({\theta}_{l_{\rm{n}}},r_{l_{\rm{n}}}\right)\right),
\end{aligned}
\end{equation}
where $L$ represents the number of all path components, $\gamma\in[0,1]$ is an adjustable parameter, which can control the proportion of the two types of path components, i.e.,  $\gamma L$ is the number of far-field path components, while $(1-\gamma)L$ is the number of near-field path components. ${\alpha_{l_{\rm{f}}}}$ and ${\theta}_{l_{\rm{f}}}$ respectively represent the path gain and angle for the ${l_{\rm{f}}}$th far-field path component, and ${\bf{a}}\left({\theta}_{l_{\rm{f}}}\right)$ is the far-field array steering vector assoicated with ${\theta}_{l_{\rm{f}}}$. ${\alpha_{l_{\rm{n}}}}$, ${\theta}_{l_{\rm n}}$, and $r_{l_{\rm{n}}}$ respectively represent the path gain, angle, and distance for the ${l_{\rm{n}}}$th near-field path component. Finally, ${\bf{b}}\left({\theta}_{l_{\rm{n}}},r_{l_{\rm{n}}}\right)$ is the near-field array steering vector assoicated with ${\theta}_{l_{\rm{n}}}$ and $r_{l_{\rm{n}}}$. When $\gamma=1$, the proposed hybrid-field channel model is simplified as the standard far-field channel model, while when $\gamma=0$, the hybrid-field channel model becomes a near-field channel model. Therefore, the proposed hybrid-field channel model is a more general channel model, where the existing far-field and near-field channel models can regarded as its special cases.


However, as mentioned in Section II, the near-field path components will cause serious energy spread in the angle domain, while the far-field path components will cause serious energy leakage in the polar domain. Thus, for the hybrid-field channel consisting of both the far-field and near-field path components, the entire channel is not sparse enough neither in the angle domain nor in the polar domain. Consequently, the existing far-field and near-field channel estimation schemes cannot be directly used to accurately estimate the hybrid-field XL-MIMO channel.

\subsection{Proposed Hybrid-Field Channel Estimation}\label{S3.2}
Based on the existing OMP algorithm for far-field channel estimation, we propose a hybrid-field OMP (HF-OMP) based hybrid-field channel estimation scheme by considering the developed hybrid-field channel model. The basic idea is that, the far-field and near-field path components are respectively estimated based on different channel transform matrices.  Specifically, since there are two types of path components rather than only one in the hybrid-field channel, different CS problems should be formulated for different path components. Based on~(\ref{eq5}) and~(\ref{eq9}), the channel estimation problem in~(\ref{eq2}) can be further represented as:
\begin{equation}\label{eq11}
\begin{aligned}
{\bf{y}} &={\bf{P}}{{{\bf{h}}}}_{\rm{f}} + {\bf{P}}{{{\bf{h}}}}_{\rm{n}} + {\bf{n}}
\\&= {\bf{PF}}{{{\bf{h}}}}_{A} + {\bf{PW}}{{{\bf{h}}}}_P + {\bf{n}},
\end{aligned}
\end{equation}
where ${{{\bf{h}}}}_{\rm{f}}$ and ${{{\bf{h}}}}_{\rm{n}}$ represent the far-field and near-field path components of the XL-MIMO channel ${{{\bf{h}}}}_{\rm{hybrid\mbox{-}field}}$, respectively. ${{{\bf{h}}}}_A$ represents the far-field path components in the angle domain, while ${{{\bf{h}}}}_{P}$ represents the near-field path components in the polar domain, which are both sparse. Therefore, if ${{{\bf{h}}}}_{A}$ and ${{{\bf{h}}}}_{P}$ can be individually estimated with the reduced pilot overhead, the whole estimated channel can be directly obtained. The specific algorithm composed of three stages can be summarized in {\textbf{Algorithm 1}}.
\begin{algorithm}[htbp]
\caption{HF-OMP based hybrid-field channel estimation}
\textbf{Inputs}: ${\bf{y}}$, ${\bf{P}}$, $\bf{F}$, $\bf{W}$, $L$, $\gamma$.
\\\textbf{Initialization}:  $L_{\rm f}=\gamma$L, $L_{\rm n}=(1-\gamma)L$, ${\Omega}_{\rm{f}}={\Omega}_{\rm{n}}=\emptyset$, ${\bf{r}}={\bf{y}}$.
\\ // Estimate far-field path components in angle domain.
\\1. ${\bf{A}}_{\rm{f}} = {\bf{P}}\bf{F}$
\\2. \textbf{for} $l_{\rm{f}} = 1,2,\cdots, L_{\rm f}$ \textbf{do}
\\3. \hspace*{+3mm}${n^{*}}={\mathop{\rm{argmax}}\limits_{n=1,2,\cdots,N}}\|{\bf{A}}^{H}_{\rm{f}}(:,n){{\bf{r}}\|}^2_2$
\\4. \hspace*{+3mm}${\Omega}_{\rm{f}}={\Omega}_{{\rm{f}}}\bigcup n^{*}$
\\5. \hspace*{+3mm}${\hat{{\bf{h}}}}_A={\bf{0}}_{N\times 1}$
\\6. \hspace*{+3mm}${\hat{{\bf{h}}}}_A({\Omega}_{\rm{f}})={{{\bf{A}}_{\rm{f}}}}^{\dag}(:,\Omega_{\rm{f}}){\bf{y}}$
\\7. \hspace*{+3mm}${\bf{r}}={\bf{y}}-{\bf{A}}_{\rm{f}}{\hat{{\bf{h}}}}_A$
\\8. \textbf{end for}
\\ // Estimate near-field path components in polar domain.
\\9. ${\bf{A}}_{\rm{n}} = {\bf{P}}\bf{W}$
\\10. \textbf{for} $l_{\rm{n}} = 1,2,\cdots,L_{\rm{n}}$ \textbf{do}
\\11. \hspace*{+3mm}${n^{*}}={\mathop{\rm{argmax}}\limits_{n=1,2,\cdots,N}}\|{\bf{A}}^{H}_{\rm{n}}(:,n){{\bf{r}}\|}^2_2$
\\12. \hspace*{+3mm}${\Omega}_{\rm{n}}={\Omega}_{\rm{n}}\bigcup n^{*}$
\\13. \hspace*{+3mm}${\hat{{\bf{h}}}}_P={\bf{0}}_{S\times 1}$
\\14. \hspace*{+3mm}${\hat{{\bf{h}}}}_P({\Omega}_{\rm{n}})={{{\bf{A}}_{\rm{n}}}}^{\dag}(:,\Omega_{n}){\bf{y}}$
\\15. \hspace*{+3mm}\textbf{if} ${\Omega}_{\rm{f}}\neq\emptyset$ \textbf{then}
\\16. \hspace*{+6mm}${\bf{r}}={\bf{y}}-{\bf{A}}_{\rm{n}}{\hat{{\bf{h}}}}_P-{\bf{A}}_{\rm{f}}{\hat{{\bf{h}}}}_A$
\\17. \hspace*{+3mm}\textbf{else}
\\18. \hspace*{+6mm}${\bf{r}}={\bf{y}}-{\bf{A}}_{\rm{n}}{\hat{{\bf{h}}}}_{P}$
\\19. \hspace*{+3mm}\textbf{end}
\\20. \textbf{end for}
\\ // Obtain all path components.
\\21. ${\hat{{\bf{h}}}}={\bf{0}}_{N\times 1}$
\\22. \textbf{if} ${\Omega}_{\rm f}\neq\emptyset$ \textbf{then}
\\23. \hspace*{+3mm}${\hat{{\bf{h}}}}={\hat{{\bf{h}}}}+{\bf{F}}{\hat{{\bf{h}}}}_{A}$
\\24. \textbf{end}
\\25. \textbf{if} ${\Omega}_{\rm n}\neq\emptyset$ \textbf{then}
\\26. \hspace*{+3mm}${\hat{{\bf{h}}}}={\hat{{\bf{h}}}}+{\bf{W}}{\hat{{\bf{h}}}}_{P}$
\\27. \textbf{end}
\\\textbf{Output}: Estimated hybrid-field channel ${\hat{{\bf{h}}}}$.
\end{algorithm}

The three stages of \textbf{Algorithm 1} can be explained as follows. Let  $L_{\rm f}=\gamma$L and $L_{\rm n}=(1-\gamma)L$ respectively denote the number of non-zero elements to be found in the angle domain and in the polar domain. Let ${\Omega}_{\rm{f}}$ and ${\Omega}_{\rm{n}}$ respectively denote the support sets assoicated with the far-field and near-field path components, which are both initialized as the empty set $\emptyset$. In the first stage, the far-field path components will be estimated in the angle domain. As shown in Step 1, the far-field sensing matrix can be represented as ${\bf{A}}_{\rm{f}} = {\bf{P}}\bf{F}$. Then $L_{\rm{f}}$ iterations will be performed to find $L_{\rm{f}}$ supports associated with $L_{\rm{f}}$ far-field path components in the angle domain. For each iteration $l_{\rm{f}}$, the correlation between the sensing matrix ${\bf{A}}_{\rm{f}}$ and the residual vector ${\bf{r}}$ needs to be calculated, where the most correlative column index in ${\bf{A}}_{\rm{f}}$ with ${\bf{r}}$ is regarded as the newly found far-field support $n^*$, as shown in Step 3. Based on the updated far-field support set ${{\Omega}}_{\rm{f}}$ in Step 4, the currently estimated far-field sparse vector ${\hat{{\bf{h}}}}_{A}$ in the angle domain is obtained by using least square (LS) algorithm in Step 6. After that, the residual vector ${\bf{r}}$ is updated by removing the contribution of far-field path components that have been estimated. Finally, we can obtain the finally estimated far-field path components ${\hat{{\bf{h}}}}_{A}$ in the angle domain after $L_{\rm{f}} $ iterations.

In the second stage, the near-field path components will be estimated in the polar domain with the near-field sensing matrix ${\bf{A}}_{\rm{n}} = {\bf{P}}\bf{W}$. The process of this stage is similar to that of the first stage, with only two differences. The first difference is that, the near-field sparse vector ${\hat{{\bf{h}}}}_{P}$ in the polar domain to be estimated in Step 14 is an $S\times 1$ vector, whose dimension is generally much higher than that of the far-field sparse vector ${\hat{{\bf{h}}}}_{A}$. The second difference is that, when updating the residual vector ${\bf{r}}$, not only the contribution of the estimated near-field path components but also that of all estimated far-field path components (if there exist) should be removed.

After estimating ${\hat{{\bf{h}}}}_{A}$ and ${\hat{{\bf{h}}}}_{P}$, we can obtain the entire estimated channel $\hat{\bf{h}}$ on the basis of~(\ref{eq5}) and~(\ref{eq9}), as shown in Steps 21-27. It is noted that both the existing far-field OMP and near-field OMP algorithms can be regarded as special cases of the proposed HF-OMP algorithm by respectively setting $\gamma=1$ and $\gamma=0$.

Finally, the computational complexity of the proposed HF-OMP algorithm is analyzed as follows. The computational complexity of the first stage and the second stage can be directly obtained as the $\mathcal{O}(NM(\gamma L)^3)$ and $\mathcal{O}(SM((1-\gamma)L)^3)$ by referring to the OMP algorithm~\cite{XinyuGao_beamsqilt_TSP}. In the third stage, the computational complexity mainly comes from Step 23 and Step 26, which can be represented by $\mathcal{O}(NS)$. To sum up, the overall computational complexity of the proposed HS-OMP algorithm is $\mathcal{O}(NM(\gamma L)^3)$+$\mathcal{O}(SM((1-\gamma)L)^3)$+$\mathcal{O}(NS)$. By contrast, the computational complexity of the far-field OMP algorithm~\cite{OMP} and the near-field OMP algorithm~\cite{Mingyao} are $\mathcal{O}(NML^3)$+$\mathcal{O}(N^2)$ and $\mathcal{O}(SML^3)$+$\mathcal{O}(NS)$, respectively. 

\vspace{-3mm}

\section{Simulation Results}\label{S5}
\vspace{-1mm}
For simulations, we consider the number of BS antennas $N=512$. The wavelength is set as $\lambda=0.01$ meters, corresponding to
the $30$ GHz frequency. The number of all path components is set as $L=6$. The path gain $\alpha_l$, angle $\theta_l$ and distance $r_l$ are generated as following:  $\alpha_l\sim{\cal CN}\left(0,1\right)$,  $\theta_l\sim{\cal U}\left(-1, 1\right)$,  and $r_l \sim{\cal U}\left(10, 80\right)$ meters.  The number of all sampled grids for the polar-domain transform matrix $\bf{W}$ is set as $S=2071$, which is generated according to the method described in~\cite{Mingyao}.  The SNR is defined as $1/{\sigma}^2$. 


We compare the proposed HF-OMP based hybrid-field channel estimation scheme with the existing far-field OMP based scheme~\cite{OMP} and the near-field OMP based scheme~\cite{Mingyao}, where the number of pilots is set as $M=256$, and each element of the pilot matrix ${\bf{P}}$ is randomly selected from ${\{-\frac{1}{\sqrt{M}},+\frac{1}{\sqrt{M}}\}}$. It is noted that since the transform matrices are generated on the sampled grids, the number of non-zero elements is larger than the number of paths. Thus, in the above three schemes, there are $12L$ non-zero elements to be estimated. Moreover, we consider the classical minimum mean square error (MMSE) based scheme as the benchmark for comparison, where the number of pilots is set as $M=512$ and the pilot matrix $\bf{P}$ is set as an identity matrix.

\begin{figure}[h]
	\begin{center}
		\includegraphics[width=0.8\linewidth]{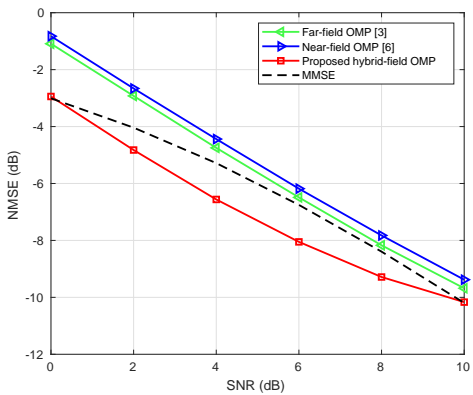}
	\end{center}
	\setlength{\abovecaptionskip}{-0.25cm}
	\caption{NMSE performance comparison against the SNR.} \label{FIG4}
	\vspace{-1mm}
\end{figure}

\begin{figure}[h]
	\begin{center}
		\includegraphics[width=0.8\linewidth]{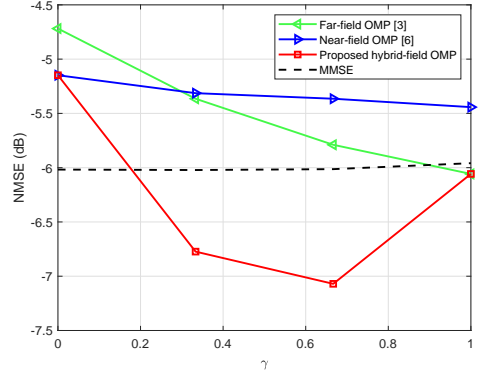}
	\end{center}
	\setlength{\abovecaptionskip}{-0.25cm}
	\caption{NMSE performance comparison against the adjustable parameter $\gamma$.} \label{FIG4}
	\vspace{-1mm}
\end{figure}


Fig. 3 shows the normalized mean square error (NMSE) performance comparison against the SNR with the adjustable parameter $\gamma=0.5$. Fig. 4 shows the NMSE performance comparison against the adjustable parameter $\gamma$, where the SNR is set as $5$ dB. We can find that if there are only near-field path components (i.e., $\gamma=0$), the proposed HF-OMP based scheme can achieve the same NMSE performance as the near-field OMP based scheme.  If there are only far-field path components (i.e., $\gamma=1$), the proposed HF-OMP based scheme can achieve the same NMSE performance as the far-field OMP based scheme.

\section{Conclusions}\label{S6}
In this paper, we have proposed a hybrid-field channel estimation scheme by accurately modeling the hybrid-field XL-MIMO channel. It is shown that the existing far-field and near-field channel estimation schemes can be regarded as special cases of the proposed hybrid-field channel estimation scheme. Simulation results show that the proposed scheme can achieve better NMSE performance with the same low pilot overhead. For future works, more advanced CS algorithms can be used to solve the hybrid-field channel estimation problem with improved performance.

\bibliography{IEEEabrv,Hybrid_Field_CE}

\begin{thebibliography}{1}
\providecommand{\url}[1]{#1}
\csname url@samestyle\endcsname
\providecommand{\newblock}{\relax}
\providecommand{\bibinfo}[2]{#2}
\providecommand{\BIBentrySTDinterwordspacing}{\spaceskip=0pt\relax}
\providecommand{\BIBentryALTinterwordstretchfactor}{4}
\providecommand{\BIBentryALTinterwordspacing}{\spaceskip=\fontdimen2\font plus
\BIBentryALTinterwordstretchfactor\fontdimen3\font minus
  \fontdimen4\font\relax}
\providecommand{\BIBforeignlanguage}[2]{{%
\expandafter\ifx\csname l@#1\endcsname\relax
\typeout{** WARNING: IEEEtran.bst: No hyphenation pattern has been}%
\typeout{** loaded for the language `#1'. Using the pattern for}%
\typeout{** the default language instead.}%
\else
\language=\csname l@#1\endcsname
\fi
#2}}
\providecommand{\BIBdecl}{\relax}
\BIBdecl

\bibitem{6G}
M.~{Giordani}, M.~{Polese}, M.~{Mezzavilla}, S.~{Rangan}, and M.~{Zorzi},
  ``Toward {6G} networks: Use cases and technologies,'' \emph{IEEE
  Commun.Mag.}, vol.~58, no.~3, pp. 55--61, Mar. 2020.

\bibitem{XLMIMO}
E.~D. {Carvalho}, A.~{Ali}, A.~{Amiri}, M.~{Angjelichinoski}, and R.~W.
  {Heath}, ``Non-stationarities in extra-large-scale massive {MIMO},''
  \emph{IEEE Wireless Commun.}, vol.~27, no.~4, pp. 74--80, Aug. 2020.

\bibitem{OMP}
J.~{Lee}, G.~{Gil}, and Y.~H. {Lee}, ``Channel estimation via orthogonal
  matching pursuit for hybrid {MIMO} systems in millimeter wave
  communications,'' \emph{{IEEE} Trans. Wireless Commun.}, vol.~64, no.~6, pp.
  2370--2386, Jun. 2016.

\bibitem{XinyuGao_beamsqilt_TSP}
X.~{Gao}, L.~{Dai}, S.~{Zhou}, A.~M. {Sayeed}, and L.~{Hanzo}, ``Wideband
  beamspace channel estimation for millimeter-wave {MIMO} systems relying on
  lens antenna arrays,'' \emph{IEEE Trans. Signal Process.}, vol.~67, no.~18,
  pp. 4809--4824, Sep. 2019.

\bibitem{My}
X.~{Wei}, C.~{Hu}, and L.~{Dai}, ``Deep learning for beamspace channel
  estimation in millimeter-wave massive {MIMO} systems,'' \emph{IEEE Trans.
  Commun.}, vol.~69, no.~1, pp. 182--193, Jan. 2021.

\bibitem{Mingyao}
M.~{Cui} and L.~{Dai}, ``Channel estimation for extremely large-scale {MIMO}:
  Far-field or near-field?'' \emph{arXiv preprint arXiv:2108.07581}, Aug. 2021.

\bibitem{JinShi}
Y.~{Han}, S.~{Jin}, C.~{Wen}, and X.~{Ma}, ``Channel estimation for extremely
  large-scale massive {MIMO} systems,'' \emph{IEEE Wireless Commun. Lett.},
  vol.~9, no.~5, pp. 633--637, May 2020.

\bibitem{Near}
X.~Yin, S.~Wang, N.~Zhang, and B.~Ai, ``Scatterer localization using
  large-scale antenna arrays based on a spherical wave-front parametric
  model,'' \emph{IEEE Trans. Wireless Commun.}, vol.~16, no.~10, pp.
  6543--6556, Jul. 2017.

\end{thebibliography}

\end{document}